\documentclass[showpacs,preprintnumbers,prd,nofootinbib,floats,amssymb,floatfix]{revtex4}
\usepackage{graphicx}
\usepackage{amsxtra}
\usepackage{amssymb}
\usepackage{amsmath}
\usepackage{bbold}
\usepackage{physics}
\usepackage{bm}
\usepackage{tensor}
\usepackage{xcolor}
\usepackage{graphicx}
\graphicspath{ {Imagenes/} }
\usepackage{hyperref}
\usepackage{amssymb}
\usepackage{amstext}
\usepackage{amsmath}
\usepackage{cleveref}
\usepackage{stackrel}
\usepackage{cancel}
\usepackage{blkarray}
\begin{document}
\title{Canonical and symplectic analysis of  the Holst action in the $G\rightarrow 0$ limit}
 \author{Victor Julián P\'erez Aquino} \email{vperezq@ifuap.buap.mx} 
 \author{Alberto Escalante}  \email{aescalan@ifuap.buap.mx}
\affiliation{Instituto de F\'isica, Benem\'erita Universidad Aut\'onoma de Puebla. \\ Apartado Postal J-48 72570, Puebla Pue., M\'exico,}
 
\begin{abstract}
By using the Dirac  and symplectic formalisms, the analysis of the Holst action in the $G \rightarrow 0$ limit is performed. We study the theory's full canonical structure and symmetries. In particular, we focus on the cases $\gamma = i$ and $\gamma \neq i$, aiming to obtain Smolin's auto-self-dual model in terms of the Ashtekar variables and a Barbero-type formulation. We then perform the symplectic analysis and  discuss the principal insight of these formalisms.

\end{abstract}
 \date{\today}
\preprint{}
\maketitle

\section{Introduction}
To better understand the dynamical structure of the gravitational field, researchers have explored new ways to study General Relativity (GR) in both configuration and phase space. In this respect, we can cite that in the 1980s, Ashtekar introduced a set of new variables, now called Ashtekar variables, which reduce the constraints of GR to a polynomial structure with a friendly structure compared to the traditional ADM constraints  \cite{Ashtekar1986PRL, Ashtekar1987PRD, Ashtekar1991NonPerturbative, ArnowittDeserMisner1962ADM, AshtekarRomanoTate1989}. With Ashtekar variables, the constraints become polynomial, but since these variables are complex, GR also becomes complex. In the 1990s, Barbero proposed a new set of real-valued variables \cite{Barbero1995}. While these variables describe GR, they lose the polynomial structure of the  Ashtekar constraints.\\
In 1996, Holst introduced an action that couples the Palatini action, via a parameter, to a new topological term called the Holst term \cite{Holst1996}. In this regard, the Hamiltonian formulation of the Palatini action is not trivial; however, when the Holst term is added and the Ashtekar variables are introduced, the analysis proceeds directly. The parameter in the Holst action, known as the Barbero-Immirzi [BI] parameter and usually written as $\gamma$, has a special role. At the Lagrangian level, $\gamma$ does not affect the equations of motion. However, at the Hamiltonian level, it changes the fundamental brackets and affects the quantum spectrum of area and volume operators in the Loop Quantum Gravity [LQG] framework \cite{Rovelli2004QuantumGravity, Thiemann2007ModernCQGR, Smolin1992}. For certain values, such as $\gamma = \pm i$, the self-dual Ashtekar formulation is recovered. For $\gamma = \pm 1$, Barbero's formulation is obtained. Whether $\gamma$ is a physical parameter is still an open question, and understanding its role is important for classical and quantum gravity research.\\
In the 1990s, Lee Smolin studied GR using Ashtekar variables \cite{Ashtekar1991NonPerturbative}. His model starts from the Ashtekar formulation and takes the $G \rightarrow 0$ limit of the curvature interaction terms, yielding a linearized self-dual formulation that describes propagating degrees of freedom. Smolin's model is valuable because it offers a different way to study GR perturbatively. In this context, GR quantization is usually done by expanding the metric in powers of $l_{planck} = \sqrt{\frac{\hbar G}{c^3}}$ around Minkowski spacetime, yielding the well-known Fierz-Pauli (FP) model. However, this approach leads to a non-renormalizable theory. If the $G \rightarrow 0$ limit is used in this model, it becomes a free field theory, but important symmetries of GR, like diffeomorphism invariance, are lost. Smolin's model stands out because it is not a free-field theory and keeps gravitational symmetries, making it a strong candidate for a perturbative theory of GR.\\
As mentioned earlier, Smolin's work used Ashtekar variables, but we believe a broader study should also include the Barbero-Immirzi parameter. Adding topological invariants or the Holst term to the Palatini action does not change the equations of motion, but it does affect the phase space, either by changing the constraints or the fundamental brackets. In this respect, it is well known that, in addition to the Holst invariant, the  Pontryagin and Euler (PE) invariants can be added. \cite{EscalanteCarbajal2011, ChandiaZanelli1998}. Like the Holst term, PE invariants do not affect the equations of motion at the Lagrangian level. However, at the Hamiltonian level, these invariants are expected to play an important role.\\
Based on the discussion above, this paper examines, from the canonical and symplectic point of view \cite{fad1, fad2, fad3, fad4}, a generalization of Smolin's model that includes the BI parameter. We expect that for $\gamma = \pm i$, Smolin's model will be recovered. For $\gamma \neq i $, we should obtain a kind of the Barbero formulation, which has not yet been described in the literature.  Then, as complementary work, we add an appendix, where we will couple the topological invariants to understand how PE invariants contribute to the canonical formulation of this GR toy with any value of the BI parameter.\\
The paper is organized as follows. In Section II, we analyze the Holst action with an arbitrary BI parameter in the $G \rightarrow 0$ limit, and then check its consistency with Smolin's results. In Section III, we present the Faddeev-Jackiw [FJ] formulation for the Holst action in the $G \rightarrow 0$ limit; we identify the complete set of constraints, and a symplectic tensor is constructed, from which we can identify the fundamental FJ brackets, confirming those obtained by  Dirac's approach.  In Section IV, we conclude. \\
\section{Holst action in the $G \rightarrow 0$ limit}
We start with the well-known Holst action
\begin{align}
    S[e, \omega] & = \int \star (e_I \wedge e_ J) \wedge R^{IJ} + \frac{1}{\gamma} \int e_I \wedge e_J \wedge R^{IJ} \nonumber \\
        & = \int d^4x \epsilon^{\mu \nu \rho \sigma} e_{\mu I} e_{\nu J} \left( \frac{1}{2}\epsilon^{IJ}_{KL}R^{KL}_{\rho \sigma} + \frac{1}{\gamma} R^{IJ}_{\rho \sigma}\right),
\end{align}
were $e^I_\alpha$ is the tetrad,  $I,J,K,...=0,1,2,3$ are the internal or group indices, while $\mu, \nu, ... = 0,1,2,3$ are spacetime indices and $\gamma$ is the so-called Barbero-Immirzi parameter. The star product is defined by $\star T^{IJ} = \frac{1}{2} \epsilon^{IJKL}T_{KL}$ and $\epsilon^{IJKL}$ is the volume 4-form associated with the internal metric $\eta_{IJ} = (-1, 0,0,0)$ and  $\epsilon^{0123} = 1$. 
The constant $G$ appears in the curvature,  and the curvature for the Lorentz connection in the $G\rightarrow0$ limit takes the form
\begin{equation}
    R^{IJ}_{\rho \sigma} = \partial_\rho \omega^{IJ}_\sigma - \partial_\sigma \omega^{IJ}_{\rho} + \cancel{ {G \omega^{IK}_{\rho} \omega ^{J}_{\sigma K} - G \omega ^{JK}_\rho \omega _{\sigma K} ^{I}}},
\end{equation}
hence, the  action will be given by 
\begin{equation}
    S[e, \omega] = \int d^4x \epsilon^{\mu \nu \rho \sigma}e_{\mu I} e_{\nu J} \left[ \frac{1}{2} \epsilon^{IJ}{}_{KL} (\partial_\rho \omega^{KL}_{\sigma} - \partial_\sigma \omega ^{KL}_\rho) + \frac{1}{\gamma} (\partial_\rho \omega_\sigma^{IJ} - \partial_\sigma \omega^{IJ}_\rho) \right].
\end{equation}
In this manner,  introducing the  ADM variables given by $N=e^0_0$ and $N^ae^i_a = e^i_0$, the Lapse and Shift functions respectively, the action is written in the following new fashion 
\begin{align}
     S[e, \omega] = & \int d^4x\eta^{abc} e_{aj}\left[  -N (\epsilon^j{}_{kl}(\partial_b\omega_c^{kl} - \partial_c \omega_b^{kl}) + \frac{2}{\gamma}(\partial_b \omega_c^{0j} - \partial_c \omega_b^{0j}))\right. \nonumber \\
    & \qquad\left.+ 2 e_{di}N^d \left( \epsilon^{ij}{}_l(\partial_b \omega_c^{l0} - \partial_c \omega_b ^{l0}) + \frac{1}{\gamma} (\partial_b \omega_c^{ij} - \partial_c \omega_b^{ij}) \right) \right.\nonumber \\
    & \qquad \left. + 2 e_{bi} \left(\epsilon^{ij}{}_l(\partial_0 \omega_c^{0l} - \partial_c \omega_0^{0l}) - \frac{1}{\gamma}(\partial_0\omega_c ^{ij} - \partial_c \omega_0^{ij}) \right) \right].
\end{align}
where Latin indices $a=1,2,3$ are the space indices, while $i= 1,2,3$ are the internal indices. We defined $ \eta^{abc} \equiv \epsilon^{0abc}$ and $\epsilon^{ijk} \equiv \epsilon^{0ijk}$.

Now we will introduce the following   variables
\begin{align}
    \omega^{ij}_\alpha &= \epsilon^{ij}{}_k\omega^k_\alpha, \\
    \omega^k_\alpha &= \gamma (A^k_\alpha + \omega_\alpha^{0k}),
\end{align}
and the action is reduced to 
\begin{align}\label{eq:lag_smolin_3+1}
    S[e^i_a, N, N^a, \omega^i_a, A^i_a, A^i_0] = & 2 \int d^4x\eta^{abc} e_{aj}\left[ -N\left( \left( \frac{\gamma^2+1}{\gamma^2} \right) F^j_{bc} - \frac{1}{\gamma}\tilde{F}^j_{bc} \right) \right.\nonumber \\ 
    & \left. + e_{ai} N^a(\epsilon^{ij}{}_l\tilde{F}^l_{bc}) - e_{bi}\epsilon^{ij}{}_k\left(\partial_0A^k_c + \partial_c\omega^{0k}_0 - \frac{1}{\gamma} \partial_c\omega^k_0\right) \right],
\end{align}
where $F^i_{ab}= \partial_a\omega^i_b - \partial_b \omega^i_a$,  $\tilde{F}^i_{ab} = \partial_a A^i_b - \partial_b A^i_a$, and $A_0^i= \omega^{0k}_0 - \frac{1}{\gamma} \omega^k_0 $. At this stage, we can take $\gamma=i$ or $\gamma=1$; however, we will do this at the end of the calculations.   By calculating the   canonical momenta of the dynamical variables given by 
\begin{align}\label{eq:momenta_smolin}
    & \pi^\alpha_k = \frac{\partial \mathcal{L}}{\partial\dot{\omega}^k_\alpha} = 0,\\
    & \tilde{\pi}^0_k = \frac{\partial \mathcal{L}}{\partial \dot{A}^k_0} = 0,\\
    & \tilde{\pi}^d_l = \frac{\partial \mathcal{L}}{\partial\dot{A}_d^l} = - \eta^{abd}e_{aj}e_{bi} \epsilon^{ij}{}_l,
\end{align}
with the identification of the fundamental Poisson brackets
\begin{align}
    \{ \omega^i_a(x), \pi^b_j(y) \} &= \delta^i_j \delta^b_a \delta(x-y),\\
    \{ A^i_a(x) , \tilde{\pi}^b_j(y)\} &=\delta^i_j \delta^b_a \delta(x-y),
\end{align}
we find that the canonical Hamiltonian is given by 
\begin{align}
    \mathcal{H}_c = \tilde{N} \epsilon_i{}^{jk}\tilde{\pi}^b_j \tilde{\pi}^c_k \left[ \left( \frac{\gamma^2+1}{\gamma^2} \right) F^i_{bc} - \frac{1}{\gamma}\tilde{F}^i_{bc} \right] + 2N^a\tilde{\pi}^b_i \tilde{F}^i_{ab} - A^k_0 \partial_c \tilde {\pi}^c_k,
\end{align}
where $\tilde{N} = \frac{N}{4e}$. It is well known that the Dirac method demands that the canonical momenta be defined for all the dynamical variables. However, if Dirac's analysis is developed on a reduced phase space, it associates canonical momenta only with those variables that appear in the action with a time derivative. In fact, we could associate their canonical momenta with the variables $\tilde{N}$, $N^a$, $A^k_0$, but these canonical variables will only extend the phase space; the canonical momenta associated with $\tilde{N}$, $N^a$, $A^k_0$ will be primary  first-class constraints, and we can work without them, obtaining the same results on the reduced phase space \cite{Al}.  \\
In this manner, because $\tilde{N}$, $N^a$, and $A^k_0$  are Lagrange multipliers, from the action, we identify the following set of  constraints 
\begin{align}
    Q & : \epsilon_i{}^{jk}\tilde{\pi}^b_j \tilde{\pi}^c_k \left[ \left( \frac{\gamma^2+1}{\gamma^2} \right) F^i_{bc} - \frac{1}{\gamma}\tilde{F}^i_{bc} \right] \approx 0, \\
    Q_a & : 2\tilde{\pi}^b_i \tilde{F}^i_{ab} \approx 0, \\
    G_k & : \partial_c \tilde {\pi}^c_k \approx 0, \\
    \phi^a_k & : \pi^a_k \approx 0.
\end{align}
where $Q, Q_a$ and $G_k$ are identified as the so-called scalar, vector and Gauss constraints, and the last one emerged from the definition of the momenta. From relevant consistency conditions, say $\dot \phi^a_l = \{\phi^a_l, H_c\}$,  we obtain the following 
\begin{equation}
 \dot \phi^a_i = \{\phi^a_i, H_c\} \rightarrow    P^a_i :  \left( \frac{\gamma^2 + 1}{\gamma^2} \right) \epsilon_i{}^{jk}[  \tilde N \tilde \pi ^b_j \partial_b \tilde \pi ^a_k + \tilde \pi ^b_j \tilde \pi ^a_k \partial_b \tilde N ] \approx 0, 
\end{equation}
where we observe that it is a relation between the Lagrange multiplier $\tilde N $ and this relation is not a constraint. However, we can observe that  the contraction of $P^a_i$ with the momenta 
\begin{equation}
    P^a_i \tilde \pi ^{ic} = \left( \frac{\gamma^2 + 1}{\gamma^2} \right) \epsilon_i{}^{jk}[  \tilde N \tilde \pi ^{ic} \tilde \pi ^b_j \partial_b \tilde \pi ^a_k + \tilde \pi^{ic} \tilde \pi ^b_j \tilde \pi ^a_k \partial_b \tilde N ],
\end{equation}
and
\begin{equation}
    P^c_i \tilde \pi ^{ia} = \left( \frac{\gamma^2 + 1}{\gamma^2} \right) \epsilon_i{}^{jk}[  \tilde N \tilde \pi ^{ia} \tilde \pi ^b_j \partial_b \tilde \pi ^c_k + \tilde \pi^{ia} \tilde \pi ^b_j \tilde \pi ^c_k \partial_b \tilde N ],
\end{equation}
then, by taking  the addition  $P^{ac} = P^a_i \tilde \pi ^{ic} + P^c_i \tilde \pi ^{ia}$, we obtain the following six secondary constraints 
\begin{equation}
    P^{ac} = \left( \frac{\gamma^2 + 1}{\gamma} \right) \epsilon^{ijk} \tilde \pi ^b_j (\tilde \pi ^c_i \partial_b \tilde \pi ^a_k + \tilde \pi^a_i \partial \tilde \pi ^c_k  ) \approx 0.
\end{equation}
To these secondary constraints  we must apply the consistency conditions, say $\dot{P}^{ac}\approx 0$,  thus 
\begin{align}
    \dot P^{ac} = & \{ P^{ac} , H_c \} = \int d^3y \{ P^{ac} (x) , \mathcal{H}_c (y) \} \nonumber \\
        = & 3 \partial_e N^e P^{ac} + N^e \partial_e P^{ac} + \partial_d N^a P^{cd} + \partial_d N^c P^{da},
\end{align}
we see that the evolution is closed; it is a linear combination of first-class constraints, and no further constraints emerge.  In this manner, the complete  set of constraints is  given by 
\begin{align}
    Q & : \epsilon_i{}^{jk}\tilde{\pi}^b_j \tilde{\pi}^c_k \left[ \left( \frac{\gamma^2+1}{\gamma^2} \right) F^i_{bc} - \frac{1}{\gamma}\tilde{F}^i_{bc} \right] \approx 0, \\
    Q_a & : 2 \tilde{\pi}^b_i \tilde{F}^i_{ab} \approx 0, \\
    P_k & : \partial_c \tilde {\pi}^c_k \approx 0, \\
    \phi^a_k & : \pi^a_k \approx 0, \\
    P^{ac} & : \left( \frac{\gamma^2 + 1}{\gamma} \right) \epsilon^{ijk} \tilde \pi ^b_j (\tilde \pi ^c_i \partial_b \tilde \pi ^a_k + \tilde \pi^a_i \partial \tilde \pi ^c_k  ) \approx 0.
\end{align}
With all the constraints in hand, we construct the matrix, say $M$, whose entries are the Poisson brackets between all the constraints. This matrix is given by  
\begin{equation}
    M= 
        \bordermatrix{    
        & \phi^a_k & P^i & P^{ac} & Q & Q_a \cr
        \phi^b_j & 0 & 0 & 0 & \{\phi^b_j,Q\} & 0 \cr
        P^j & 0 & 0 & 0 & 0 & 0 \cr
        P^{bd} & 0 & 0 & 0 & 0 & 0 \cr
        Q & \{Q,\phi^a_k\} & 0 & 0 &  \{Q,Q\} & 0 \cr
        Q_b & 0 & 0 & 0 & 0 & 0 \cr
                    },
    \end{equation}
where the non-trivial brackets are expressed by 
\begin{align}
\label{braq}
    \{\phi^a_l(x),Q(y) \} & = 2 \epsilon_l {}^{jk} \left( \frac{\gamma^2+1}{\gamma^2} \right) \tilde \pi^a_j (y) \tilde \pi^c_k(y)  {\partial_c} \delta^3(x-y), \\
    \{Q(x),Q(y) \} & = \frac{4}{\gamma} \epsilon_i{}^{jk}\epsilon_k{}^{qr} \left[ U^i_{bc}(x) \tilde \pi^d_q (y) \tilde \pi ^c_r (y) \tilde \pi^b_j (x) {\partial_d} \delta^3(x-y) - U_{qde}(y) \tilde \pi ^b_j (x) \tilde \pi ^{di}(x) \tilde \pi ^e_r (y) {\partial_b} \delta^3(x-y) \right],
\end{align}
with  $U^i_{ab}(x)=\left[ \left( \frac{\gamma^2+ 1}{\gamma^2} \right) F^i_{ab}(x) - \frac{1}{\gamma^2} \tilde F^i_{ab} (x) \right]$. Hence, the set of constraints can be classified  as follows
\textbf{First class constraints:}
    \begin{align*}
        & P_i : \partial_c \tilde{\pi}^c_i,\\
        & Q_a : 2\tilde{\pi}^b_i \tilde{F}^i_{ab}, \\
        & P^{ac}: \left( \frac{\gamma^2+1}{\gamma} \right) \epsilon^{ijk}\tilde{\pi}^b_j(\tilde{\pi}^c_i \partial_b\tilde{\pi}^a_k +  \tilde{\pi}^a_i\partial_b\tilde{\pi}^c_k).
    \end{align*}
\textbf{Second class constraints:}
    \begin{align*}
        & Q : \epsilon_i{}^{jk} \tilde{\pi}^b_j \tilde{\pi}^c_k \left[ \left( \frac{\gamma^2 + 1}{\gamma^2} \right) F^i_{bc}  - \frac{1}{\gamma} \tilde{F}^i_{bc} \right], \\ 
        & \phi^a_k : \pi^a_k,
    \end{align*} 
whit the identification of the constraints, we can perform he counting of degrees of freedom as follows; there are  $36$ canonical variables in the phase space,  12 first class constraints and 10 second class constraints, hence  
\begin{equation*}
    \text{DOF} = \frac{1}{2} \left( \left(^\text{phase spase}_\text{variables} \right) - 2 \left(^\text{first class}_\text{constraints}\right)  - \left(^\text{second class}_\text{constraint}\right)  \right) 
\end{equation*}

\begin{equation}
    \text{DOF} = \frac{1}{2} (36 - 2(12)-10) = 1.
\end{equation}
This model  has one degree of freedom, and the scalar constraint is classified as  a second-class constraint; therefore, a gauge symmetry of the theory is lost that is generated by the usual scalar constraint of covariant theories. However, this is not the whole story; let us analyze our results as follows. First,  we  observe that if  $\gamma=  i$, then our results are reduced to 
\begin{equation}
    \mathcal{H}_c =  i \tilde N \epsilon_o{}^{jk} \tilde \pi^b_j \tilde \pi^c_k \tilde F^i_{bc} + 2 N^a \tilde \pi^b_i \tilde F^i_{ab} - A_0^k \partial_c \tilde \pi^c_k,
\end{equation}
and the constraints will be 
\begin{align}
    & P_i : \partial_c \tilde{\pi}^c_i \approx 0,\\
    & Q_a : 2\tilde{\pi}^b_i \tilde{F}^i_{ab} \approx 0, \\
    & Q : \epsilon_i{}^{jk} \tilde{\pi}^b_j \tilde{\pi}^c_k \left[ i \tilde{F}^i_{bc} \right], \approx 0
\end{align}
which are all of first class. This matches what Smolin reported in \cite{Smolin1992GNewtonZero}, and because of the complex constraints, we must add reality conditions. In this scenario, the counting of degrees of freedom is now $DOF = \frac{1}{2}[18-2(7)] = 2$ as expected. Therefore, for  $\gamma=  i$ we are dealing with a system where the self-dual connection is propagating on an anti-self-dual background, and the Gauss constraint is  the generator of three copies of $U(1)^3$. On the other hand, if we take $\gamma\neq i$, we must solve the second-class constraints, or we can construct the corresponding Dirac's brackets. Let us take the second way. In fact,  the matrix whose entries are the Poisson brackets between the second-class constraints is given by 
\begin{equation}
C_{\alpha\beta}=
\begin{pmatrix}
0 & \{\phi_i^a,Q\}\\[2mm]
-\{\phi_j^b,Q\} & \{Q,Q\}
\end{pmatrix},
\end{equation}
where these entries are given in the equations (\ref{braq}). Hence, for constructing the Dirac's brackets we need calculate the  inverse of  $C_{\alpha \beta}$, where it is defined by  
\begin{equation}
    \int d^3y C_{\alpha \beta} (x,y) (C^{\beta \gamma})^{-1}(y,z) = \delta^\gamma_\alpha \delta^3(x-z).
\end{equation}
 However, we can observe the following,  if we write $C^{-1}$  such as 
\begin{equation}
    C C^{-1} = 
    \begin{pmatrix}
    0 & \{\phi_i^a,Q\}\\[2mm]
    -\{\phi_j^b,Q\} & \{Q,Q\}
    \end{pmatrix}
    \begin{pmatrix}
    X & Y\\[2mm]
    Z & W
    \end{pmatrix}
    = I_{10}=
    \begin{pmatrix}
    I_9 & 0\\[2mm]
    0 & I
    \end{pmatrix},
\end{equation}
from this  we  observe that 
\begin{equation}
    \{\phi^a_i,Q\}Z = I_9,
\end{equation}
hence, $Z$ must be an operator that takes nine functions and projects just one. Therefore, this equation has infinitely many solutions,  since there are infinitely many combinations of the tensor's structure given in $\{\phi_j^b,Q\}$ with  nine components that can satisfy this single scalar divergence given in (\ref{braq}). In this manner, this observation allows us to conclude that  the solution for  avoiding this problem is to take   $\omega_a^i =0$. In this scenario, the constraints are reduced to 
 \begin{align}
 \label{conso}
    & P_i : \partial_c \tilde{\pi}^c_i \approx 0,\\
    & Q_a : 2\tilde{\pi}^b_i \tilde{F}^i_{ab} \approx 0, \\
    & Q : \epsilon_i{}^{jk} \tilde{\pi}^b_j \tilde{\pi}^c_k \left[ \frac{1}{\gamma} \tilde{F}^i_{bc} \right] \approx 0.
\end{align}
 However, as consequence of   $\omega^i_a=0$, is that the connection takes the form $A^i_a = - \gamma \omega^{0k}_a= -\gamma K_a^i$. So these are not the usual Ashtekar variables, and the constraints given in (\ref{conso}) do not describe Smolin's model. In fact, 
 in this scenery,    the intrinsic curvature $^{(3)}R^c{}_{dab} = 0$, however, there is the extrinsic curvature. Hence,  if a spatial slice is perfectly flat (like a smooth sheet of rubber, without bumps or mountains), you can stretch it in one direction (say, the $x$-axis) while compressing it in another (say, the $y$-axis). During this process, the sheet remains intrinsically flat ($\omega_a^i = 0$), yet it undergoes a drastic change in shape. The extrinsic curvature $K_a^i$ encodes precisely these rates of deformation. Mathematically, volume-preserving deformations in three spatial dimensions are described by a symmetric, traceless matrix with exactly two independent modes. Thus, $\gamma =i$ and $\gamma\neq i$ predict different scenarios: for the former, the theory is reduced to Smolin's purpose; for the latter, it could be like the Barbero formulation, only that the principal geometric ingredient will be the extrinsic curvature. 
  \section{The Faddeev-Jackiw analysis} 
 In this section, we develop the FJ analysis, a powerful tool for analyzing gauge systems \cite{ fad1,  fad2, fad3, fad4}.
The method is a symplectic approach in which all relevant information of the theory is obtained through an invertible symplectic tensor constructed from symplectic variables identified as degrees of freedom. Because the theory is singular, constraints arise. The FJ framework treats all constraints equally, so it is not necessary to classify them as primary, secondary, first class, or second class as in Dirac’s method [14,15]. Once the symplectic tensor is obtained, its components correspond to the FJ generalized brackets. In the end, Dirac’s brackets and FJ brackets coincide.  Thus, we will use this formalism to study the critical case, when $\gamma \neq i$,  for comparing the results with those developed in the previous section. \\
From the action  (\ref{eq:lag_smolin_3+1}) we identify the following symplectic lagrangian
\begin{eqnarray}
\nonumber
 \mathcal{L}^{(0)}&=&\pi_{i}^{a} \dot{A}_{a}^{i}+A_{0}^{i} \partial_{a} \pi_{i}^{a}+N \varepsilon_{i}^{j k} \pi_{j}^{b} \pi_{k}^{c}\left[\left(\frac{\gamma^{2}+1}{\gamma^{2}}\right)\left(\partial_{b} \omega_{c}^{i}-\partial_{c} \omega_{b}^{i}\right)-\frac{1}{\gamma}\left(\partial_{b} A_{c}^{i}-\partial_{c} A_{b}^{i}\right)\right] \\ 
& +&N^{b} \pi_{i}^{c}\left(\partial_{b} A_{c}^{i}-\partial_{c} A_{b}^{i}\right),
\end{eqnarray}
then, the symplectic Lagrangian has the form $ \mathcal{L}^{(0)}=\pi_{i}^{a} \dot{A}_{a}^{i}- V^{(0)}$, where we identify the following symplectic potential  
\begin{eqnarray}
\nonumber 
V^{(0)} &=& -A_{0}^{i} \partial_{a} \pi_{i}^{a}-N \varepsilon_{i}^{j k} \pi_{j}^{b} \pi_{k}^{c}\left[\left(\frac{\gamma^{2}+1}{\gamma^{2}}\right)\left(\partial_{b} \omega_{c}^{i}+\partial_{c} \omega_{b}^{i}\right)-\frac{1}{\gamma}\left(\partial_{b} A_{c}^{i}-\partial_{c} A_{b}^{i}\right)\right] \\ 
&-&N^{b} \pi_{i}^{c}\left(\partial_{b} A_{c}^{i}-\partial_{c} A_{b}^{i}\right), 
\end{eqnarray} 
thus the symplectic equations of motion are given by 
\begin{equation}
\label{eqmov}
f_{\alpha \beta}^{(0)} \dot{\xi}^\beta= \frac{\partial V^{(0)}}{\partial \xi^ \alpha}, 
\end{equation}
where $f_{\alpha \beta}^{(0)}$ is called the symplectic matrix and it is given by 
$$f^{(0)}_{\alpha\beta}(x,y) = \frac{\delta a_\beta(y)}{\delta \xi^\alpha(x)} - \frac{\delta a_\alpha(x)}{\delta \xi^\beta(y)}.$$
From the symplectic Lagrangian we identify the following symplectic variables  $\xi^{\alpha}=\left(A_{0}^{i}, A_{a}^{i}, \omega_{a}^{i}, N, N^{a}, \pi^a_i
\right)$,  and the one-forms  $a_{\alpha}=\left(0, \pi_{i}^{a}, 0,0,0,0 \right)$.  It is worth commenting on a formal distinction between the Dirac and Faddeev-Jackiw frameworks regarding the lapse $N$ and shift $N^a$. In the Dirac analysis, these functions serve strictly as external Lagrange multipliers. In contrast, within the symplectic Faddeev-Jackiw framework, all variables appearing in the symplectic potential—including multipliers—are formally treated on equal footing as symplectic coordinates $\xi^\alpha$ within the symplectic potential $V^{(0)}$. The absence of time derivatives for these multipliers results in vanishing rows in the initial symplectic matrix $f_{\alpha\beta}^{(0)}$, whose zero modes systematically generate the constraints of the theory without requiring manual consistency checks, as we shall see below. Hence,  by using the symplectic variables, we calculate the symplectic matrix  given by 
\begin{equation}
f_{\alpha\beta}^{(0)} = \begin{pmatrix} 0 & 0 & 0 & 0 & 0 & 0 \\ 
0 & 0 & 0 & 0 & 0 & -\delta_i^j \delta_a^b \\ 
0 & 0 & 0 & 0 & 0 & 0 \\ 
0 & 0 & 0 & 0 & 0 & 0 \\ 
0 & 0 & 0 & 0 & 0 & 0 \\
 0 & \delta_j^i \delta_b^a & 0 & 0 & 0 & 0 \end{pmatrix} \delta^3(x-y), 
\end{equation}
from this  symplectic matrix we observe that it is singular, then  we identify the following null vectors 
\begin{eqnarray}
\nonumber 
\mathcal{V}_1 &=& (v^{A_{0}^{i}},0, 0, 0, 0, 0), \\ \nonumber
\mathcal{V}_2 &=& (0, 0, v^{\omega_a^i}, 0, 0,0), \\ \nonumber
\mathcal{V}_3 &=& (0, 0, 0, v^{N}, 0, 0), \\ 
\mathcal{V}_4 &=& (0, 0, 0, 0, v^{N^{a}}, 0), 
\end{eqnarray}
 from these null vectors, we obtain  by means of  \cite{fad1, fad2, fad3, fad4}
 \begin{equation}
\Omega_k= \int dx^3 \mathcal{V}_k^{\alpha}\frac{\partial}{\partial \xi^{\alpha}(x)}\int V^{(0)}(\xi(y))dy^3, 
 \end{equation}
 the following  constraints
 \begin{eqnarray}
\Omega_{1i} &=& \partial_{a} \pi_{i}^{a}, \\ 
\Omega_2 &=& \varepsilon_{i}^{j k} \pi_{j}^{b} \pi_{k}^{c}\left[\left(\frac{\gamma^{2}+1}{\gamma^{2}}\right)\left(\partial_{b} \omega_{c}^{i}-\partial_{c} \omega_{b}^{i}\right)-\frac{1}{\gamma}\left(\partial_{b} A_{c}^{i}-\partial_{c} A_{b}^{i}\right)\right], \\
\Omega_{3d} &=&\pi_{i}^{b} (\partial_d A_b^i - \partial_b A_d^i), \\
\Omega_4^{eb} &=& \pi_{l}^{g}\left(\pi^{e j} \varepsilon_{j}{^{l k}} \partial_{g} \pi_{k}^{b}+\pi^{b j} \varepsilon_{j}{^{l k}} \partial_{g} \pi_{k}^{e}\right).
\end{eqnarray}
At this stage, we observe that the null vector $\mathcal{V}_2$ indicates constraints, and these are given by $\Omega_4^{eb}$, which correspond to those found in the Dirac framework. In fact, the symplectic zero modes of the initial matrix directly yield the complete constraint hypersurface, reproducing the secondary constraint $\Omega_4^{eb}$ without requiring the explicit iterative consistency equations, say  ($\dot{\phi} \approx 0$),  characteristic of Dirac's algorithm, thereby highlighting the computational efficiency of the symplectic approach. Furthermore, we demand consistency conditions on these constraints,  similar to  Dirac's approach; this is 
\begin{equation}
\label{consis}
\dot{\Omega}_k=\frac{\partial \Omega}{\partial \xi^\alpha}\dot{\xi}^{\alpha}=0,
\end{equation}
from the combination between (\ref{eqmov}) and (\ref{consis}), we obtain the following generalized system 
\begin{eqnarray}
f^{(1)}_{\alpha \beta} \dot{\xi}^{\beta}= Z_k(\xi), 
\end{eqnarray}
where 
\begin{equation}
f^{(1)}_{\alpha\beta} = \begin{pmatrix}  f^{(0)}_{\alpha\beta}& \\
\frac{\partial \Omega_k}{\partial \xi^\alpha}
 \end{pmatrix},
\end{equation}
 and 
 \begin{equation}
 Z_k(\xi)=  \begin{pmatrix}  \frac{\partial V^ {(0)}}{\partial \xi^\alpha}& \\
0
 \end{pmatrix},
\end{equation}
 hence we calculate $f^{(1)}_{\alpha\beta}$ and it is given by 
 \begin{equation}
 \label{m1}
f_{\alpha\beta}^{(1)} = \begin{pmatrix} 0 & 0 & 0 & 0 & 0 & 0 \\ 
0 & 0 & 0 & 0 & 0 & -\delta_i^j \delta_a^b \\ 
0 & 0 & 0 & 0 & 0 & 0 \\ 
0 & 0 & 0 & 0 & 0 & 0 \\ 
0 & 0 & 0 & 0 & 0 & 0 \\
 0 & \delta_j^i \delta_b^a & 0 & 0 & 0 & 0\\
 0&0&0&0&0&  \delta_i^j \partial_b \delta^3(x-y)\\
 0&\frac{\delta \Omega_2(x)}{\delta A_{j}^{b}(y)}& \frac{\delta \Omega_2(x)}{\delta \omega_{a}^{i}(y)} &0&0&\frac{\delta \Omega_2(x)}{\delta \pi_{i}^{a}(y)}\\
 0& \frac{\delta \Omega_{3d}(x)}{\delta A_{a}^{i}(y)}&0&0&0&\frac{\delta \Omega_{3d}(x)}{\delta \pi_{i}^{a}(y)} \\
 0&0&0&0&0& \frac{\delta \Omega_4^{eb}(x)}{\delta \pi_i^a(y)} 
  \end{pmatrix} \delta^3(x-y), 
\end{equation}
where 
\begin{eqnarray}
\label{variat}
\frac{\delta \Omega_2(x)}{\delta A_{j}^{b}(y)} &=& -\frac{2}{\gamma} \varepsilon_j^{\ l k} \pi_l^d \pi_k^b(x) \partial_d \delta^3(x-y), \nonumber \\
\frac{\delta \Omega_2(x)}{\delta \omega_{a}^{i}(y)} &=& 2 \left(\frac{\gamma^2+1}{\gamma^2}\right) \varepsilon_i^{\ j k} \pi_j^b(x) \pi_k^a(x) \partial_b\delta^3(x-y), \nonumber \\ 
\frac{\delta \Omega_2(x)}{\delta \pi_{i}^{a}(y)} &=& 2 \varepsilon_i^{\ i k} \pi_k^c(x) \Big[ \left(\frac{\gamma^2+1}{\gamma^2}\right) (\partial_a \omega_c^i - \partial_c \omega_a^i) - \frac{1}{\gamma}(\partial_a A_c^i - \partial_c A_a^i)\Big] \delta^3(x-y), \nonumber \\ 
\frac{\delta \Omega_{3d}(x)}{\delta A_{a}^{i}(y)} &=& \pi_i^a(x) \partial_d \delta^3(x-y) - \pi_i^b(x) \delta_d^a \partial_b \delta^3(x-y), \nonumber \\
\frac{\delta \Omega_{3d}(x)}{\delta \pi_{i}^{a}(y)} &=& (\partial_d A_a^i(x) - \partial_aA_d^i(x)) \delta^3(x-y), \nonumber \\
\frac{\delta \Omega_4^{eb}(x)}{\delta \pi_i^a(y)} &=& \Big[ \varepsilon_{j}^{\ i k} \left[ \pi^{ej}(x) \partial_a \pi_k^b(x) + \pi^{bj}(x) \partial_a \pi_k^e(x) \right] + \pi_l^g(x) \varepsilon^{i l k} \left[ \delta_a^e \partial_g \pi_k^b(x) + \delta_a^b \partial_g \pi_k^e(x) \right] \Big]\delta^3(x-y) \nonumber  \\ 
&+&   \pi_l^g(x) \varepsilon_{j}^{\ l i} \left[ \pi^{ej}(x) \delta_a^b + \pi^{bj}(x) \delta_a^e \right] \partial_g\delta^3(x-y),
\end{eqnarray}
we can observe that the matrix (\ref{m1}) has the following null vectors given by 
\begin{eqnarray}
\mathcal{V}_1 &=& (0, \partial_b v^{\Omega_1}, 0, 0, 0, 0, v^{\Omega_1}, 0, 0, 0),\nonumber \\
\mathcal{V}_2 &=& \left(0, v^{\Omega_3} \frac{\delta \Omega_{3d}}{\delta \pi_i^a}, 0,  0, 0,  -v^{\Omega_3} \frac{\delta \Omega_{3d}}{\delta A_a^i},  0,  0,  v^{\Omega_4},  0 \right), \nonumber \\
\mathcal{V}_3 &=& \left(0,  v^{\Omega_4} \frac{\delta \Omega_4^{eb}}{\delta \pi_i^a},  0,  0,  0,  0,  0,  0, 0,  v^{\Omega_4} \right), 
\end{eqnarray}
 and  $Z_k(\xi)$ takes the form
 \begin{equation}
 Z_k(\xi) = \begin{pmatrix}  -\Omega_{1i} \\  -\frac{2}{\gamma} \partial_b \left( N \varepsilon_{i}^{j k} \pi_{j}^{b} \pi_{k}^{a} \right) + \partial_b (N^b \pi_i^a) - \partial_b (N^a \pi_i^b) \\  2 \left( \frac{\gamma^2+1}{\gamma^2} \right) \partial_b \left( N \varepsilon_{i}^{j k} \pi_{j}^{b} \pi_{k}^{a} \right) \\  -\Omega_2 \\  -\Omega_{3a} \\  \partial_a A_0^i - 2 N \varepsilon{_{l}}^{i k} \pi_{k}^{c} U_{a c}^{l} - N^b (\partial_b A_a^i - \partial_a A_b^i) \\  0 \\ 0 \\ 0 \\ 0 \end{pmatrix}
 \end{equation}
 with $U_{ac}^i = \left(\frac{\gamma^2+1}{\gamma^2}\right)(\partial_a \omega_c^i - \partial_c \omega_a^i) - \frac{1}{\gamma}(\partial_a A_c^i - \partial_c A_a^i)$. 
 Thus, from the contraction of the null vectors with $Z_k(\xi)$, we obtain that  $\mathcal{V}_1^k Z_k=0$, therefore, no more constraints emerge.  \\
 Because there are no more constraints, we will add to the symplectic Lagrangian the constraints  $\Big(\Omega_{1i}, \Omega_2, \Omega_{3d}, \Omega_4^{eb}\Big) $ with their corresponding Lagrange multipliers $\Big( \rho^i, \alpha, \sigma^a, \beta_{eb}\Big)$, respectively; then we obtain 
\begin{eqnarray}
\nonumber
\mathcal{L}^{(2)}&= &\pi_{i}^{a} \dot{A}_{a}^{i}+\dot{\rho}^i \partial_{a} \pi_{i}^{a}+ \dot{\beta}_{eb} \pi_{l}^{g}\left(\pi^{e j} \varepsilon_{j}{^{l k}} \partial_{g} \pi_{k}^{b}+\pi^{b j} \varepsilon_{j}{^{l k}} \partial_{k} \pi_{k}^{e}\right) \\  \nonumber 
&+&\dot{\alpha}  \varepsilon_{i}^{j k} \pi_{j}^{b} \pi_{k}^{c}\left[\left(\frac{\gamma^{2}+1}{\gamma^{2}}\right)\left(\partial_{b} \omega_{c}^{i}-\partial_{c} \omega_{b}^{i}\right)-\frac{1}{\gamma}\left(\partial_{b} A_{c}^{i}-\partial_{c} A_{b}^{i}\right)\right] \\  
&+& \dot{\sigma}^a \pi_{i}^{b} (\partial_aA_b^i- \partial_bA_a^i)- V^{(1)},
\end{eqnarray}
where $V^{(1)}= V^{(0)}|_{\Omega_k}=0$. In this manner, we can identify from the symplectic Lagrangian the new  symplectic variables   $\xi^{\alpha}=\left( \rho^{i}, A_{a}^{i}, \omega_{a}^{i}, \alpha, \sigma^a , \pi^a_i, \beta_{eb} 
\right)$, and the one-forms 
$a_{\alpha}=\Big(\partial_{a} \pi_{i}^{a},  \pi_{i}^{a}, 0,   \varepsilon_{i}^{j k} \pi_{j}^{b} \pi_{k}^{c}\left[\left(\frac{\gamma^{2}+1}{\gamma^{2}}\right)\left(\partial_{b} \omega_{c}^{i}-\partial_{c} \omega_{b}^{i}\right)-\frac{1}{\gamma}\left(\partial_{b} A_{c}^{i}-\partial_{c} A_{b}^{i}\right)\right],  \pi_{i}^{b} (\partial_aA_b^i- \partial_bA_a^i), 0,  
\pi_{l}^{g}\left(\pi^{e j} \varepsilon_{j}{^{l k}} \partial_{g} \pi_{k}^{b}+\pi^{b j} \varepsilon_{j}{^{l k}} \partial_{k} \pi_{k}^{e}\right) \Big)$. With these symplectic variables, we calculate the new symplectic matrix
 given by 
\begin{equation}
\label{M1}
f^{(2)}_{\alpha\beta}(x,y) = \begin{pmatrix} 0 & 0 & 0 & 0 & 0 & - \frac{\delta \Omega_1(x)}{\delta \pi(y)} & 0 \\ 0 & 0 & 0 & \frac{\delta \Omega_2(y)}{\delta A(x)} & \frac{\delta \Omega_3(y)}{\delta A(x)} & - \delta_j^i \delta_b^a \delta^3(x-y) & 0 \\ 0 & 0 & 0 & \frac{\delta \Omega_2(y)}{\delta \omega(x)} & 0 & 0 & 0 \\ 0 & - \frac{\delta \Omega_2(x)}{\delta A(y)} & - \frac{\delta \Omega_2(x)}{\delta \omega(y)} & 0 & 0 & - \frac{\delta \Omega_2(x)}{\delta \pi(y)} & 0 \\ 0 & - \frac{\delta \Omega_3(x)}{\delta A(y)} & 0 & 0 & 0 & - \frac{\delta \Omega_3(x)}{\delta \pi(y)} & 0 \\ \frac{\delta \Omega_1(y)}{\delta \pi(x)} & \delta_i^j \delta_a^b \delta^3(x-y) & 0 & \frac{\delta \Omega_2(y)}{\delta \pi(x)} & \frac{\delta \Omega_3(y)}{\delta \pi(x)} & 0 & \frac{\delta \Omega_4(y)}{\delta \pi(x)} \\ 0 & 0 & 0 & 0 & 0 & - \frac{\delta \Omega_4(x)}{\delta \pi(y)} & 0 \end{pmatrix} \delta^3(x-y),
\end{equation}
where the variations of the constraints are given in (\ref{variat}). This matrix is singular, in fact, there are the following null vectors 
\begin{eqnarray}
\nonumber 
 \mathcal{V}_1^{(1)}& = &\{ v^{\Omega_1}, - \partial_a v^{\Omega_1},  0,  0,  0,  0, 0 \}, \nonumber \\
 \mathcal{V}_2^{(1)} &=& \{0,  - \frac{\delta \Omega_{3b}}{\delta \pi_i^a} v^{\Omega_3},  0,  0 , v^{\Omega_3},  \frac{\delta \Omega_{3b}}{\delta A_a^i} v^{\Omega_3},  0 \}, \nonumber \\
 \mathcal{V}_3^{(1)} &=& \{ 0,  0,  v^\omega,   0,  0,   0, 0 \}, 
\end{eqnarray}
we can observe that the null vectors $\mathcal{V}_1^{(1)}$ and $ \mathcal{V}_2^{(1)}$ are the generators of $U(1)$ gauge symmetry and spatial diffeormorphisms respectively. In fact, in the FJ formalism, a null vector $\mathcal{V}^\alpha$ of the symplectic matrix $f_{\alpha\beta}$ satisfies $f_{\alpha\beta} \mathcal{V}^\beta = 0$. This algebraic property means that the components of the null vector define the infinitesimal variations (or flow) of the symplectic variables $\xi^\beta$ along the gauge orbits: $\delta \xi^\beta = \epsilon \, \mathcal{V}^\beta$, where $\epsilon$ is an infinitesimal local parameter  \cite{fad1, fad2, fad3, fad4}.  Thus, by taking to $v^{\Omega_1} \equiv \epsilon^i(x)$ as an arbitrary local parameter, and applying the variation $\delta \xi^\alpha = \mathcal{V}_1^{(1)}$, we find the non-zero infinitesimal transformations for the dynamical fields
\begin{eqnarray}
\delta \rho^i &=& \epsilon^i, \nonumber \\
\delta A_a^i &=& -\partial_a \epsilon^i, \nonumber  \\
\delta \omega_a^i &=& 0,  \quad  \delta \alpha = 0,  \quad  \delta \sigma^a = 0,  \quad  \delta \pi_i^a = 0, \quad  \delta \beta_{eb} = 0, \nonumber 
\end{eqnarray}
we observe that the transformation $\delta A_a^i = -\partial_a \epsilon^i$ is the standard infinitesimal $U(1)^3$  gauge transformation of the connection 1-form, and  $\rho^i$ acts as the Lagrange multiplier  associated with the  Gauss constraint. Moreover, for the vector $ \mathcal{V}_2^{(1)}$, we use the variations given by  $\frac{\delta \Omega_{3d}(x)}{\delta A_a^i(y)} = \left[ \pi_i^a(x) \partial_d - \pi_i^b(x) \delta_d^a \partial_b \right] \delta^3(x-y), \quad \frac{\delta \Omega_{3d}(x)}{\delta \pi_i^a(y)} = (\partial_d A_a^i - \partial_a A_d^i)\delta^3(x-y)$. Hence, we compute the induced variations $\delta \xi^\alpha = \mathcal{V}_2^{(1)}$ on the fundamental phase space variables
\begin{eqnarray}
\nonumber
\delta \sigma^a &=& \xi^a, \nonumber \\
\delta A_a^i &=& - \frac{\delta \Omega_{3b}}{\delta \pi_i^a} \xi^b = - (\partial_b A_a^i - \partial_a A_b^i)\xi^b, \nonumber \\
\delta \pi_i^a &=& \frac{\delta \Omega_{3b}}{\delta A_a^i} \xi^b = \left( \pi_i^a \partial_b \xi^b - \pi_i^b \partial_b \xi^a + \xi^b \partial_b \pi_i^a \right) = \partial_b (\xi^b \pi_i^a) - \partial_b (\xi^a \pi_i^b), \nonumber
\end{eqnarray}
rewriting the variation of the connection using vector calculus identities, $-\xi^b (\partial_b A_a^i - \partial_a A_b^i) = \xi^b \partial_b A_a^i + A_b^i \partial_a \xi^b - \partial_a(\xi^b A_b^i)$, we recognize the first two terms as the spatial components of the Lie derivative of a 1-form along the vector field $\xi^a$ ($\mathcal{L}_\xi A_a^i$). Similarly, the transformation of $\pi_i^a$ matches the exact weight-one tensor-density Lie derivative transformation under spatial coordinate reparametrizations.\\
On the other hand, the null vector   $\mathcal{V}_3^{(1)}$ generates an inconsistency, because this sector, labeled by requiring  $\frac{\delta \Omega_2(x)}{\delta \omega_a^i(y)} v^\omega(y) = 0$, can not be removed and will always be a null sector of the matrix (\ref{M1}). 
However, since the symplectic matrix (\ref{m1}) shows no additional constraints, the theory has a gauge symmetry, as expected. To obtain a symplectic tensor, we need to fix the gauge. We can start by fixing the usual gauge $A_0^i=0, N=0, N^a=0$; however, we will find that the theory is still singular because the  $\omega_a^i$ sector is present. Hence, to obtain a physical theory free of inconsistencies, we take $A_0^i=0, N=0, N^a=0$ and the addition of   $\omega_a^i=0$ as gauge fixing. The inconsistency found by the Dirac method is reflected in the construction of the Dirac bracket: the matrix between these second-class constraints does not have an inverse. In the context of FJ, however, the same inconsistency is clearer to see using null vectors. By adding these gauge fixings, the symplectic Lagrangian  takes the form
\begin{eqnarray}
\nonumber
\mathcal{L}^{(3)}&=&\pi_{i}^{a}  \dot{A}_{a}^{i}  +\dot{\rho}^i \Big(\partial_{a} \pi_{i}^{a} + \kappa_i \Big)+ \dot{\beta}_{eb} \Big( \pi_{l}^{g}\left(\pi^{e j} \varepsilon_{j}{^{l k}} \partial_{g} \pi_{k}^{b}+\pi^{b j} \varepsilon_{j}{^{l k}} \partial_{k} \pi_{k}^{e}\right) + \lambda^{eb}\Big)\\  \nonumber 
&+&\dot{\alpha} \Big(  \varepsilon_{i}^{j k} \pi_{j}^{b} \pi_{k}^{c}\left[\left(\frac{\gamma^{2}+1}{\gamma^{2}}\right)\left(\partial_{b} \omega_{c}^{i}-\partial_{c} \omega_{b}^{i}\right)-\frac{1}{\gamma}\left(\partial_{b} A_{c}^{i}-\partial_{c} A_{b}^{i}\right)\right] +\nu\Big)  \\  
&+& \dot{\sigma}^a \Big( \pi_{i}^{b} (\partial_a A_b^i- \partial_b A_a^i ) + \chi_a \Big) + \dot{ \mu}^a_i \omega^a_i.
\end{eqnarray} 
We observe at this stage that $\omega_a^i$ is not yet eliminated from the symplectic Lagrangian; we will do so after performing the variations, then set $\omega_a^i=0$.  In this manner,  the new symplectic variable are  identified by  $\xi^{\alpha}=\left( \rho^{i}, A_{a}^{i}, \omega_{a}^{i}, \alpha, \sigma^a , \pi^a_i, \beta_{eb} , \lambda^{eb}, \nu, \chi_a, \mu^a_i, \kappa_i
\right)$, and the one-forms \\
$a_{\alpha}=\Big(\partial_{a} \pi_{i}^{a} +\kappa_i,  \pi_{i}^{a}, 0,   \varepsilon_{i}^{j k} \pi_{j}^{b} \pi_{k}^{c}\left[\left(\frac{\gamma^{2}+1}{\gamma^{2}}\right)\left(\partial_{b} \omega_{c}^{i}-\partial_{c} \omega_{b}^{i}\right)-\frac{1}{\gamma}\left(\partial_{b} A_{c}^{i}-\partial_{c} A_{b}^{i}  \right) \right]+ \nu,  \pi_{i}^{b} (\partial_aA_b^i- \partial_bA_a^i) + \chi_a,  0,  
\pi_{l}^{g}\left(\pi^{e j} \varepsilon_{j}{^{l k}} \partial_{g} \pi_{k}^{b}+\pi^{b j} \varepsilon_{j}{^{l k}} \partial_{k} \pi_{k}^{e} +\lambda^{eb}\right), 0, 0, 0 ,\omega^i_a, 0   \Big)$. Hence, the symplectic matrix is given by 
\begin{equation}
f_{\alpha\beta}^{(3)}(x,y) = \begin{pmatrix} 0 & 0 & 0 & 0 & 0 & -\delta_i^j \partial_b & 0 & 0 & 0 & 0 & 0 & -\delta_i^j \\ 0 & 0 & 0 & \frac{\delta\Omega_2}{\delta A_a^i} & \frac{\delta\Omega_{3b}}{\delta A_a^i} & -\delta_i^j\delta_a^b & 0 & 0 & 0 & 0 & 0 & 0 \\ 0 & 0 & 0 & \frac{\delta\Omega_2}{\delta \omega_a^i} & 0 & 0 & 0 & 0 & 0 & 0 & \delta_i^j\delta_a^b & 0 \\ 0 & -\frac{\delta\Omega_2}{\delta A_j^b} & -\frac{\delta\Omega_2}{\delta \omega_j^b} & 0 & 0 & -\frac{\delta\Omega_2}{\delta \pi_j^b} & 0 & 0 & -1 & 0 & 0 & 0 \\ 0 & -\frac{\delta\Omega_{3a}}{\delta A_j^b} & 0 & 0 & 0 & -\frac{\delta\Omega_{3a}}{\delta \pi_j^b} & 0 & 0 & 0 & -\delta_a^b & 0 & 0 \\ \delta_j^i \partial_a & \delta_j^i\delta_b^a & 0 & \frac{\delta\Omega_2}{\delta \pi_i^a} & \frac{\delta\Omega_{3b}}{\delta \pi_i^a} & 0 & \frac{\delta\Omega_4^{cd}}{\delta \pi_i^a} & 0 & 0 & 0 & 0 & 0 \\ 0 & 0 & 0 & 0 & 0 & -\frac{\delta\Omega_4^{eb}}{\delta \pi_j^b} & 0 & -\delta_{eb}^{cd} & 0 & 0 & 0 & 0 \\ 0 & 0 & 0 & 0 & 0 & 0 & \delta_{cd}^{eb} & 0 & 0 & 0 & 0 & 0 \\ 0 & 0 & 0 & 1 & 0 & 0 & 0 & 0 & 0 & 0 & 0 & 0 \\ 0 & 0 & 0 & 0 & \delta_b^a & 0 & 0 & 0 & 0 & 0 & 0 & 0 \\ 0 & 0 & -\delta_j^i\delta_b^a & 0 & 0 & 0 & 0 & 0 & 0 & 0 & 0 & 0 \\ \delta_j^i & 0 & 0 & 0 & 0 & 0 & 0 & 0 & 0 & 0 & 0 & 0 \end{pmatrix} \delta^3(x-y).
\end{equation}
After long calculations, we see that this symplectic matrix is non-singular; thus, it is a symplectic tensor. The inverse is given by 
\begin{align*}
(f^{(3)})^{-1}(x,y) = \left( \begin{array}{cccccccc} 0 & 0 & 0 & 0 & 0 & 0 & 0 & 0  \\  
 0 & 0 & 0 & 0 & 0 & \delta_i^j \delta_a^b & 0 & \frac{\delta \Omega_4^{cd}(y)}{\delta \pi_i^a(x)} \\ 
 0 & 0 & 0 & 0 & 0 & 0 & 0 & 0 \\   
  0 & 0 & 0 & 0 & 0 & 0 & 0 & 0  \\ 
  0 & 0 & 0 & 0 & 0 & 0 & 0 & 0  \\ 
  0 & -\delta_i^j \delta_b^a & 0 & 0 & 0 & 0 & 0 & 0  \\
  0 & 0 & 0 & 0 & 0 & 0 & 0 & \delta_{(e}^c \delta_{b)}^d \\ 
  0 & -\frac{\delta \Omega_4^{eb}(x)}{\delta \pi_j^b(y)} & 0 & 0 & 0 & 0 & -\delta_{(c}^e \delta_{d)}^b & 0  \\ 
  0 & -\frac{\delta \Omega_2(x)}{\delta \pi_j^b(y)} & 0 & -1 & 0 & -\frac{\delta \Omega_2(x)}{\delta A_j^b(y)} & 0 & \{\Omega_2(x), \Omega_4^{cd}(y)\}  \\  
0 & -\frac{\delta \Omega_{3a}(x)}{\delta \pi_j^b(y)} & 0 & 0 & -\delta_a^b & -\frac{\delta \Omega_{3a}(x)}{\delta A_j^b(y)} & 0 & \{\Omega_{3a}(x), \Omega_4^{cd}(y)\}  \\ 
0 & 0 & \delta_i^j \delta_b^a & 0 & 0 & 0 & 0 & 0 \\
-\delta_i^j & -\frac{\delta \Omega_{1i}(x)}{\delta \pi_j^b(y)} & 0 & 0 & 0 & 0 & 0 & 0 \\ 
 \end{array} \right. \\
\left. \begin{array}{cccc} 
  0 & 0 & 0 & \delta_i^j \\
   \frac{\delta \Omega_2(y)}{\delta \pi_i^a(x)} & \frac{\delta \Omega_{3b}(y)}{\delta \pi_i^a(x)} & 0 & \frac{\delta \Omega_{1j}(y)}{\delta \pi_i^a(x)} \\ 
     0 & 0 & -\delta_i^j \delta_a^b & 0 \\ 
     1 & 0 & 0 & 0 \\ 
      0 & \delta_a^b & 0 & 0 \\
         \frac{\delta \Omega_2(y)}{\delta A_a^i(x)} & \frac{\delta \Omega_{3b}(y)}{\delta A_a^i(x)} & 0 & 0 \\ 
           0 & 0 & 0 & 0 \\ 
  \{\Omega_4^{eb}(x), \Omega_2(y)\} & \{\Omega_4^{eb}(x), \Omega_{3b}(y)\} & 0 & 0 \\
  \{\Omega_2(x), \Omega_2(y)\} & \{\Omega_2(x), \Omega_{3b}(y)\} & \frac{\delta \Omega_2(x)}{\delta \omega_j^b(y)} & \{\Omega_2(x), \Omega_{1j}(y)\} \\ 
    \{\Omega_{3a}(x), \Omega_2(y)\} &  \{\Omega_{3a}(x), \Omega_{3b}(y)\} &0 & \{\Omega_{3a}(x), \Omega_{1j}(y)\} \\ 
     -\frac{\delta \Omega_2(y)}{\delta \omega_a^i(x)} & 0 & 0 & 0 \\ 
      \{\Omega_{1i}(x), \Omega_2(y)\} & \{\Omega_{1i}(x), \Omega_{3b}(y)\} & 0 & 0
  \end{array}\right) \delta^3(x-y),
\end{align*}
where the constraints has the form 
 \begin{eqnarray}
\Omega_{1l} &=& \partial_{a} \pi_{l}^{a}, \nonumber  \\ 
\Omega_2 &=& \varepsilon_{i}^{j k} \pi_{j}^{b} \pi_{k}^{c}\left[\frac{1}{\gamma}\left(\partial_{b} A_{c}^{i}-\partial_{c} A_{b}^{i}\right)\right], \nonumber \\
\Omega_{3d} &=&\pi_{i}^{b} (\partial_d A_b^i - \partial_b A_d^i), 
\end{eqnarray}
and its algebra is closed  given by 
\begin{eqnarray}
&\{\Omega_{1i}(x), \Omega_{1j}(y)\} &= 0, \nonumber \\
&\{\Omega_{1i}(x), \Omega_4^{eb}(y)\} &= 0,  \nonumber \\
&\{\Omega_4^{ab}(x), \Omega_4^{cd}(y)\} & = 0,  \nonumber \\
&\{\Omega_{3c}(x), \Omega_{3d}(y)\}& = \Omega_{3c}(y) \partial_d \delta^3(x-y) + \Omega_{3d}(x) \partial_c \delta^3(x-y), \nonumber \\
&\{\Omega_{3d}(x), \Omega_{1i}(y)\}& = 0, \nonumber \\
&\{\Omega_{3d}(x), \Omega_2(y)\}& = \Omega_2(y) \partial_d \delta^3(x-y) - \partial_{d} \left( \Omega_2(y) \delta^3(x-y) \right) + \Omega_2(x) \partial_d \delta^3(x-y), \nonumber \\
&\{\Omega_{1i}(x), \Omega_2(y)\}& = 0, \nonumber \\
&\{\Omega_2(x), \Omega_2(y)\} &= \frac{8}{\gamma^2}  \pi_i^b \pi_i^c \Omega_{3c}(x) \partial_b \delta^3(x-y).
\end{eqnarray}
Now, in the FJ formalism, the fundamental brackets are given by  $\{ \xi^\alpha(x), \xi^\beta(y) \}_D = (f^{(2)})^{-1 \alpha\beta}(x,y)$. In particular, we observe that  $\{A_a^i(x), \pi_j^b(y)\}_D = \delta_a^b \delta_i^j \delta^3(x-y)$ just like it  was obtained from the Dirac framework. The counting of degrees of freedom can be performed on the reduced phase space as follows: starting with the fundamental canonical pairs $(A_a^i, \pi_i^a)$ comprising 18 phase-space variables, and accounting for the seven independent constraints along with their seven corresponding gauge-fixing conditions, the number of physical degrees of freedom is given by
\begin{equation}
\text{DoF} = \frac{1}{2}(18 - 14) = 2,
\end{equation}
confirming that the theory consistently propagates two physical degrees of freedom.
\section{Conclusions}
In this paper, the canonical and symplectic analysis of the Holst action in the $G \rightarrow 0$ limit was performed.  From the canonical analysis, and for arbitrary $\gamma$,  we identified the complete set of constraints of the theory. In this respect, we observed that for $\gamma=i$ the theory reduces to that reported by Smolin; all constraints are first class, and the theory propagates two degrees of freedom on an anti-self-dual background. On the other hand, for $\gamma \neq i$, there is an inconsistency in the construction of Dirac's brackets; then the fixing $\omega_a^i=0$ was necessary for obtaining a consistent model. The theory describes the propagation of two degrees of freedom; however, in this case, the intrinsic curvature vanishes, and only the extrinsic curvature remains.\\
Furthermore, we performed a symplectic analysis. To view the inconsistency from another perspective, we used the FJ approach. In this case, we obtained the full set of constraints and found the inconsistency in the symplectic matrix. The symplectic matrix was singular, and the standard fixing gauge could not remove the singularity; we then had to fix $\omega_a^i=0$, after which we obtained a symplectic tensor. From the entries of the symplectic tensor, the FJ brackets are identified.  Finally, a comparative assessment of both formalisms reveals complementary strengths: Dirac's method relies on systematically classifying primary and secondary constraints into first and second classes, then constructing Dirac brackets, while the Faddeev-Jackiw approach treats all constraints democratically via symplectic zero modes, making gauge-sector identification transparent. Both methodologies ultimately converge on identical physical brackets and confirm the propagation of two physical degrees of freedom for the self-dual $\gamma = i$ limit and a consistent extrinsic-curvature-driven formulation for $\gamma \neq i$.\\
We also added an appendix. In the appendix, we added the topological Pontryagin and Euler terms to the Holst action. We identified the complete set of constraints and observed that, for $\gamma = i$, we obtained the linearized version of self-dual gravity coupled to the self-dual PE. We observed that the topological terms contribute to the canonical momenta; however, the structure is similar to the Ashtekar constraints. 
\section{Appendix}
In this appendix, we  will add the topological terms of Euler and Pontryagin to our previous results. The action we started with is 
\begin{equation}\label{invariantes}
    S[\omega] = \underbrace{\int R_{IJ} \wedge R^{IJ}}_{\text{Pontryagin}} - \frac{1}{\gamma} \underbrace{\int \star R_{IJ} \wedge R^{IJ}}_{\text{Euler}}.
\end{equation}
Before perform the analysis, we will write the the action in a more suitable fashion. For this,  we introduce the following fields   
\begin{equation}
    F_{IJ} = R_{IJ} - \frac{1}{\gamma} \star R_{IJ},
\end{equation}
\begin{equation}
    \tilde{F}_{IJ}= F_{IJ} + \frac{1}{\gamma} \star F_{IJ},
\end{equation}
These  leads to the relations 
\begin{equation}
\label{rela}
    R_{IJ} = \frac{\gamma^2}{\gamma^2+1} \tilde{F}_{IJ}, \hspace{0.6cm} \tilde{F}_{IJ} = \frac{\gamma^2 + 1}{\gamma^2} R_{IJ},
\end{equation}
Hence the action  takes the form
\begin{align}
    S[\omega] & = \int R_{IJ} \wedge R^{IJ} - \frac{1}{\gamma} \int \star R_{IJ} \wedge R^{IJ} = \int \left( R_{IJ} - \frac{1}{\gamma}\star R_{IJ} \right) \wedge R^{IJ} \nonumber \\ \nonumber
    & = \int F_{IJ} \wedge R^{IJ}\\ 
    & = \frac{\gamma^2}{\gamma^2 + 1}\int F_{IJ} \wedge \tilde{F}^{IJ}.
\end{align}
Hence, by performing the $3+1$ decomposition  we obtain
\begin{align*}
    S[\omega]  = \frac{\gamma^2}{\gamma^2+1} \int d^4x \eta^{abc}[2 \tilde{F}^{0j}_{ab}F_{0j0c} + \tilde{F}^{ij}_{ab} F_{ij0c}],
\end{align*}
where we have defined $\eta^{abc} \equiv \epsilon^{0abc} $. From the relations \eqref{rela} we obtain 
\begin{align*}
    \tilde{F}^{0j}_{ab} = \frac{\gamma^2 }{\gamma ^2 + 1} R^{0j}_{ab}, \\
    \tilde{F}^{ij}_{ab} = \frac{\gamma ^2 }{\gamma^2 + 1} R ^{ij}_{ab},
\end{align*}
in the following, by considering in  the curvature (the $G \rightarrow 0 $ limit ), this is 
\begin{equation}
    R^{IJ}_{\mu \nu} = \partial_\mu \omega_ \nu^{IJ} - \partial_\nu \omega_ \mu^{IJ} + \cancel{ G \omega^{IK}_\mu \omega^J_{\nu K} -  G \omega^{JK}_\mu \omega^I_{\nu K}}.
\end{equation}
Then, by using these definitions, and introducing the variables  
\begin{align}
    \omega^{ij}_a = \epsilon^{ij}{}_k \omega^k_a,\\
    \omega ^k _a = \gamma (A^k _a + \omega ^{0k}_a),
\end{align}
we find that  the  curvature and the auxiliary fields can be rewritten as
\begin{align}
    & R^{0i}_{ab } = \frac{1}{\gamma}(\partial_a \omega^i_b - \partial_b \omega^i _a ) - (\partial_a A^i_b - \partial_b A^i _a),\\
    & R^{ij} _{ab} = \epsilon^{ij}{}_k(\partial_a \omega^k_b - \partial_b \omega^k_a),\\
    & F^{ij}_{\mu \nu} = \epsilon^{ij}_k \left[ \left( \frac{\gamma^2 +1 }{\gamma^2} \right) (\partial_\mu \omega^k_\nu - \partial_\nu \omega^k _{\mu}) - \frac{1}{\gamma} (\partial_\mu A^k_\nu -\partial_\nu A^k_\mu 
    ) \right],\\
    & F^{0i}_{\mu \nu} = - (\partial_\mu A^i_\nu - \partial_\nu A^i _\mu).
\end{align}
Hence, the action principle is finally written as 
\begin{align*}
    S[A, \omega] = 2 \int d^4x\eta^{abc} \left[ (\partial_a A^j _b - \partial_bA^j_a)(\partial_0A_{cj} - \partial_c A_{0j}) + \left( \frac{\gamma^2+1}{\gamma^2} \right) (\partial_a \omega_{bk} - \partial_b\omega_{ak})(\partial_0\omega^k_c - \partial_c \omega^k_0) \right].
\end{align*}
In this manner, we will resume the canonical analysis. The canonical Hamiltonian will be given by 
\begin{align*}
    \mathcal{H}_c & = \partial_a \tilde{\pi}^{a}_iA_{0}^i + \partial_a\pi^a_i\omega^i_0,
\end{align*}
and the complete set of constraints are 
\begin{align}
    & \phi_i^d : \pi_i^d - 2 \eta^{abd} \left( \frac{\gamma^2 +1}{\gamma^2} \right) (\partial_a \omega_{bi} - \partial_b \omega_{ai}) \approx 0, \\
    & \tilde \phi_i ^d : \tilde \pi_i^d - 2 \eta^{abd} (\partial_a A_{bi} - \partial_b A_{ai}) \approx 0, \\
    & \psi_i : \partial_a \pi^a_i \approx 0,\\
    & \tilde \psi_i : \partial_a \tilde \pi^a_i \approx 0,
\end{align}
where $\tilde\pi_i^a$ and $\pi_a^i$ are the canonical momenta of  $A_a^i$ and $\omega_a^i$ respectively. These constraints are first class, and reducible: $\partial_d \phi_i^d = \psi_i$ and $\partial_d \tilde \phi_i^d = \tilde \psi_i$.  Counting the degrees of freedom shows that the theory lacks physical degrees of freedom. In addition, if we take $\gamma=i$, then we obtain the linearized version of the self-dual representation of the PE invariants.\\
On the other hand, if we add the topological invariants to the Holst action, we obtain the following canonical Hamiltonian
\begin{align}
    \mathcal{H}_c & = \tilde N \epsilon_i{}^{jk} (\tilde \pi ^b_j - \eta^{deb} \tilde F_{dej} ) (\tilde \pi ^c_k - \eta ^{fgc} \tilde F _{fgk}) \left[ \left( \frac{\gamma^2+1}{\gamma^2}  \right) F^i_{bc} - \frac{1}{\gamma} \tilde F^i_{bc}  \right] \nonumber \\
    & - N^a \tilde F^i_{ab} (\tilde \pi^b_i - \eta ^{edb} \tilde F_ {edi}) + \tilde \pi ^c_k \partial_c A^k_0 + \pi^c_k \partial_c \omega ^k _0
\end{align}
where 
\begin{align}
    & \pi^d_i = \left( \frac{\gamma^2+1}{\gamma^2} \right) \eta^{abd} F_{abi}, \\
    & \tilde \pi ^d_l = \frac{1}{2} \eta^{abd} e_{ja} e_{ib} \epsilon^{ij}{}_l + \eta^{abd}\tilde F_{abl},
\end{align}
where $F_{ab}^i =\partial_aA_b^i- \partial_b A_a^i$ and $\tilde F_{ab}^i =\partial_a \tilde A_b^i- \partial_b \tilde A_a^i$. The constraints are given by 
\begin{align}
    & H :  \epsilon_i{}^{jk} (\tilde \pi ^b_j - \eta^{deb} \tilde F_{dej} ) (\tilde \pi ^c_k - \eta ^{fgc} \tilde F _{fgk}) \left[ \left( \frac{\gamma^2+1}{\gamma^2}  \right) F^i_{bc} - \frac{1}{\gamma} \tilde F^i_{bc}  \right] \approx 0, \\
    & V_a : \tilde F^i_{ab} (\tilde \pi^b_i - \eta ^{edb} \tilde F_ {edi}) \approx 0, \\
    & G_i : \partial_c \pi ^c_i \approx 0,\\
    & \tilde G_i : \partial_c \tilde \pi^c_i \approx 0,
\end{align}
if we take the particular case $\gamma=i$, then all our results are reduced to 
\begin{align}
    & \tilde G_i : \partial_c \tilde \pi ^c_i \approx 0, \\
    & V_a : \tilde F_{ab}^i (\tilde \pi^b_i - \eta^{bed} \tilde F_{edi}) \approx 0, \\
    & H : \epsilon_i{}^{jk} (\tilde \pi ^b_j - \eta^{deb} \tilde F_{dej} ) (\tilde \pi ^c_k - \eta ^{fgc} \tilde F _{fgk})  \tilde F^i_{bc}  \approx 0, \\
\end{align}
where that structure corresponds to the linearized ($G \rightarrow 0$ limit) form of self-dual gravity coupled to the self-dual representation of the topological invariants reported in \cite{meche}. Again, for $\gamma=i$, the system describes the propagation of two degrees of freedom. Furthermore, for arbitrary real $\gamma$,  we must fix $\omega_a^i=0$. Only the extrinsic curvature will remain, and the contribution of the topological terms will be given in the structure of the momentum terms in the constraints. \\

\end{document}